\nonstopmode

\documentclass[prb,showpacs,twocolumn]{revtex4}
\usepackage{graphicx}
\usepackage{dcolumn}
\usepackage{bm}
\usepackage{mathrsfs}

\begin{document}
\title{Electromagnetically induced Talbot effect}
\author{Jianming Wen,$^{1,2}$\footnote{Email Address: jianming.wen@gmail.com} Shengwang Du,$^3$ Huanyang Chen,$^4$ and Min Xiao$^{1,2}$\footnote{Email Address:
mxiao@uark.edu}}
\affiliation{$^1$National Laboratory of Solid State Microstructures and Department of Physics, Nanjing University, Nanjing 210093, China\\
$^2$Department of Physics, University of Arkansas, Fayetteville, Arkansas 72701, USA\\
$^3$ Department of Physics, The Hong Kong University of Science and
Technology, Clear Water Bay, Kowloon, Hong Kong, China\\
$^4$School of Physical Science and Technology, Soochow University,
Suzhou 215006, China}
\date{today}

\begin{abstract} By modulating transmission function of a weak probe field via a strong
control standing wave, an electromagnetically induced grating can be
created in the probe channel. Such a nonmaterial grating may lead to
self-imaging of ultra-cold atoms or molecules in the Fresnel
near-field regime. This work may offer a nondestructive and lensless
way to image ultra-cold atoms or molecules.
\end{abstract}

\pacs{42.50.Gy, 42.25.Fx, 42.65.An, 42.25.Bs}
\maketitle

Optical imaging offers a powerful diagnostic methodology for a
variety of experiments involving ultra-cold atoms and molecules. Two
methods, on/off-resonant absorption imaging, are often chosen to
image the atomic or molecular cloud. On-resonant absorption imaging
\cite{ketterle} is predominant, despite its limited dynamic range
and recoil heating. Although off-resonant phase-imaging techniques
\cite{dobrek} allow nondestructive and quantitative imaging of
Bose-Einstein condensates (BECs), traditional approaches usually
require precisely aligned phase plates or interferometers. In this
paper, we propose another type of lensless imaging scheme,
electromagnetically induced Talbot effect (EITE) or
electromagnetically induced self-imaging (EISI), for ultra-cold
atoms and molecules. This work might broaden variety applications in
imaging techniques and be useful for atom lithography \cite{atom} as
well.

The proposed scheme to conduct lensless imaging of ultra-cold atoms
and molecules relies on periodically manipulating transmission and
dispersion profiles of a weak probe field under the condition of
electromagnetically induced transparency (EIT).\cite{EIT1,EIT2} The
basic idea is to utilize a strong control standing wave to modify
the optical response of the medium to the weak probe field, i.e.,
electromagnetically induced grating.\cite{EIG} Such an induced
nonmaterial grating thus leads to self-images of atoms and molecules
in the Fresnel diffraction region. Conventional Talbot effect
\cite{talbot1,talbot2,case} relates the self-imaging of periodic objects
without using any optical components. Images observed in the Fresnel
near-field regime are repeated along or backward along the
illumination direction depending on whether the light is transmitted
or reflected. Recent renewed interest in this remarkable phenomenon,
Talbot effect, is motivated by the progress made on atomic
waves,\cite{pritchard} BECs,\cite{phillips} quantum
carpets,\cite{berry}, Gauss sums,\cite{schleich} x-ray phase
imaging,\cite{david} nonclassical light,\cite{wen1} and second
harmonic generation.\cite{wen2,wen22}

To be specific, we consider a medium of length $L$ consisting of an
ensemble of closed three-level ultra-cold atoms (or molecules) in
the $\Lambda$ configuration, with two metastable lower states, as
shown in Fig. 1(a). Two ground states $|b\rangle$ and $|c\rangle$
are coupled to the excited state $|a\rangle$ via a strong control
field of angular frequency $\omega_c$ near resonance on the
$|c\rangle\rightarrow|a\rangle$ transition with a detuning
$\Delta_c=\omega_c-\omega_{ac}$, and a weak probe field of angular
frequency $\omega_p$ close to resonance on the
$|b\rangle\rightarrow|a\rangle$ transition with a detuning
$\Delta_p=\omega_p-\omega_{ab}$. The control light consists of two
fields which are symmetrically displaced with respect to $z$ and
whose intersection generates a standing wave within the medium, see
Fig. 1(b) and the snapshot [Fig. 1(c)]. To introduce the notations,
we denote the amplitude of the probe field as $E_p$ and its Rabi
frequency as $\Omega_p$. For simplicity, hereafter we assume that
two cw control fields share the same Rabi frequency $\Omega_c$.
$\gamma_a$ and $\gamma_{bc}$ are the decay rate of excited state
$|a\rangle$ and dephasing rate between two ground states $|b\rangle$
and $|c\rangle$, respectively. Initially, all the population is
distributed in the ground state $|b\rangle$.
\begin{figure}[tbp]
\centerline{\includegraphics[scale=0.5]{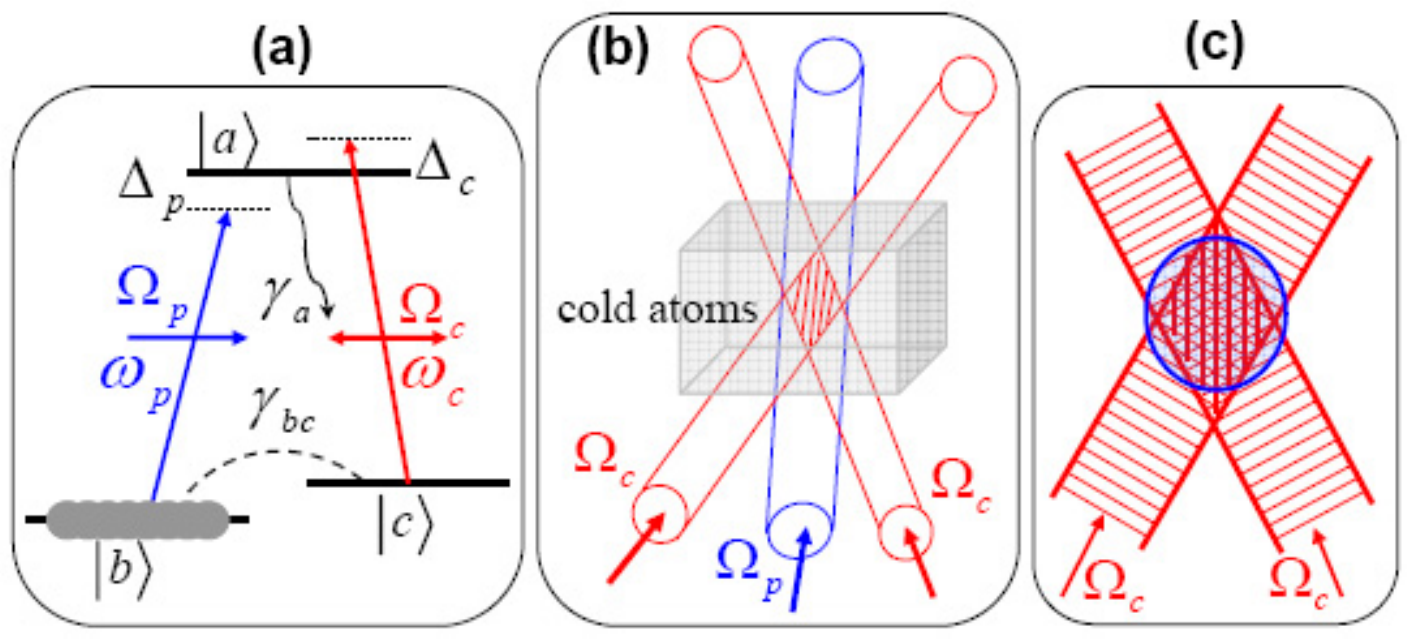}} \caption{(Color
online) (a) A closed three-level $\Lambda$-type atomic system for
EITE. (b) Configuration of forming an EIG. (c) Snapshot of the EIG
in (b).}\label{fig:Fig1}
\end{figure}

The linear susceptibility of the system at the probe frequency is
now
\begin{eqnarray}
\chi=\frac{N|\mu|^2}{2\hbar\epsilon_0}\frac{\Delta_2+i\gamma_{bc}}{|\Omega_c|^2\cos^2\big(\frac{{\pi}x}{a}\big)-(\Delta_p+i\gamma_a)(\Delta_2+
i\gamma_{bc})},
\end{eqnarray}
where $\Delta_2=\Delta_p-\Delta_c$ is the two-photon detuning and
$a$ is the spatial period of the standing wave along the $x$
direction perpendicular to the propagation direction $z$. In
principle, $a$ can be made arbitrarily larger than the wavelength
$\lambda_p$ of the probe field by varying the angle between the two
wave vectors of two control beams. The propagation dynamics of the
probe field within the cloud obeys the Maxwell's equation and its
transmission at the output surface is
\begin{eqnarray}
E_p(x,L)=E_p(x,0)e^{-\frac{k_p\chi''}{2}L}e^{i\frac{k_p\chi'}{2}L},
\end{eqnarray}
where $\chi=\chi'+i\chi''$ and $E_p(x,0)$ is the input probe
profile. For simplicity, the probe field is assumed to be a plane
wave. Figure 2(a) displays two typical profiles of the probe field
at the output surface as $\Delta_p=\Delta_2=0$. At the transverse
locations around the nodes (of the standing wave), the probe beam is
absorbed according to the usual Beer law because the control field
intensity there is very weak. In contrast, at the transverse
locations around the antinodes, the probe is absorbed much less due
to the EIT effect. This leads to a periodic amplitude modulation
across the probe profile, a phenomenon reminiscent of the amplitude
grating. If the probe field is detuned off the resonance, a phase
modulation will be also introduced to its output profile. Figure
2(b) shows, over several periods, the transmission function
$|E_p(L,x)|$ (upper curve) and phase $\Phi$ (lower curve) as
$\Delta_p=2\gamma_a$ and $\Delta_c=0$. At nodes, the probe field
experiences a rapid phase change in contrast to Fig. 2(a) where no
phase modulation is introduced.
\begin{figure}[tbp]
\includegraphics[scale=0.35]{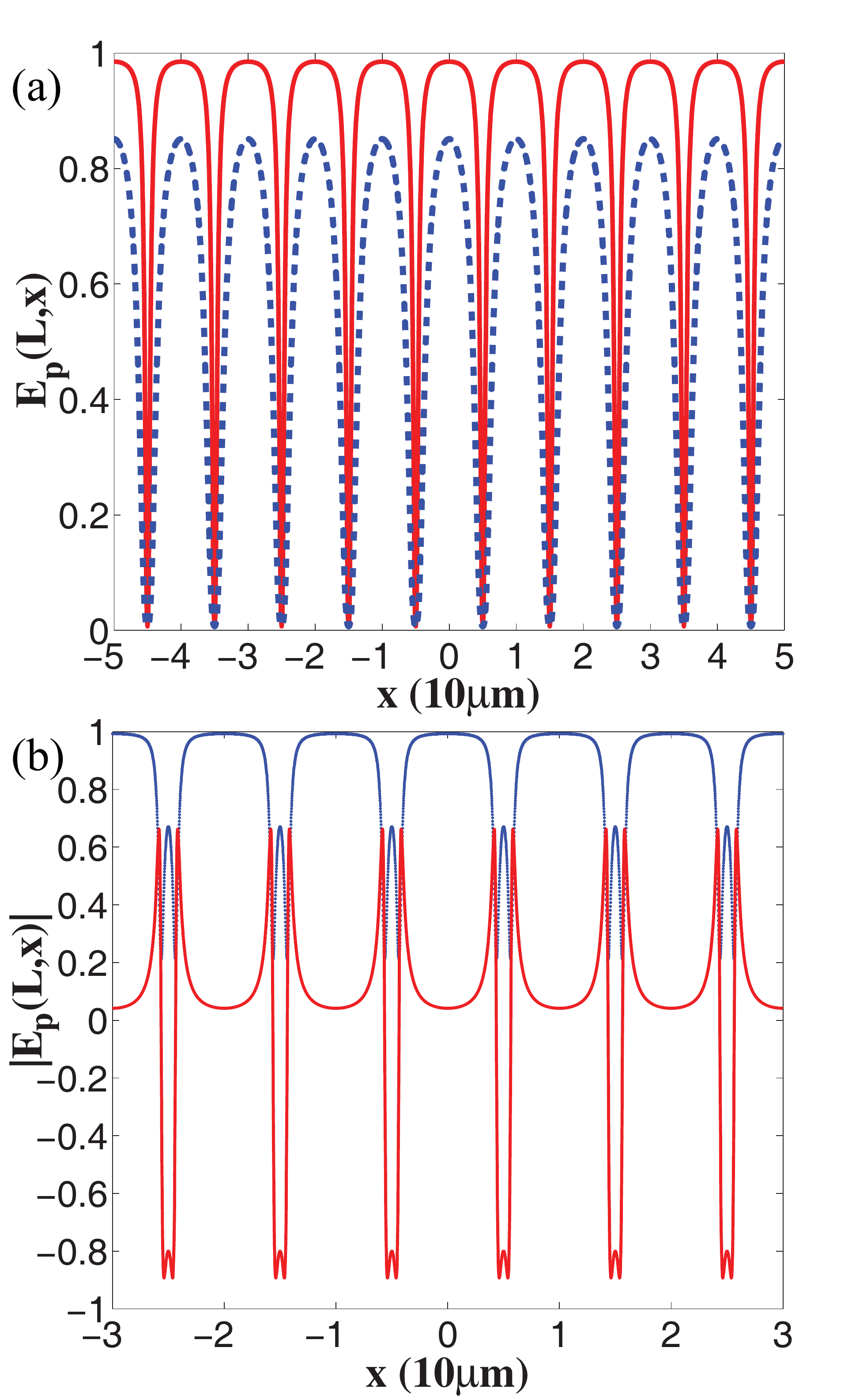}
\caption{(Color online) The output profile of the probe field,
$E_p(L,x)$, as a function of $x$. (a) An amplitude grating: optical
depth $=20$, $\Omega_c=10\gamma_a$ (upper curve) and optical depth
$=4$, $\Omega_c=3\gamma_a$ (lower curve). Other parameters are
$\Delta_p=\Delta_2=0$, $\gamma_{bc}=0.3\gamma_a$, and $a=10$ $\mu$m.
(b) A hybrid grating: optical depth $=8$, $\Delta_p=2\gamma_a$,
$\Delta_c=0$, $\Omega_c=10\gamma_a$, and
$\gamma_{bc}=0.3\gamma_a$.}\label{fig:Fig1}
\end{figure}

Using the Fresnel-Kirchhoff diffraction integral, the output probe
field $\mathscr{E}_p$ at a distance $Z$ from the output surface of
the medium is proportional to
\begin{eqnarray}
\mathscr{E}_p(X,Z)\propto\int^{\infty}_{-\infty}E_p(x,L)e^{ik_p(Z+\frac{x^2}{2Z}-\frac{xX}{Z}+\frac{X^2}{2Z})}dx,
\end{eqnarray}
where $x$ and $X$ are the coordinates in the \textit{object} and
observation planes, respectively. From Eq. (3) the diffraction
amplitude $\mathscr{E}_p(X,Z)$ is determined by $E_p(x,L)$. For
conventional Talbot effect, if $E_p$ is a one-dimensional periodic
function, self-imaging of the object then repeats at every Talbot
plane. In fact, because of the periodicity exhibited in Eq. (1) it
may be recast into Fourier series,
\begin{eqnarray}
E_p(x,L)=\sum^{+\infty}_{n=-\infty}C_ne^{i2\pi{n}\frac{x}{a}},
\end{eqnarray}
where $C_n$ is the coefficient of the $n$th harmonic. Its detailed
format may be obtained by setting $y=e^{i2\pi\frac{x}{a}}$,
representing $C_n$ in terms of the contour integral around the unit
circle, and calculating the residue at 0. By substituting Eq. (4)
into (3) and completing the integral, we recover the traditional
Talbot effect,
\begin{eqnarray}
\mathscr{E}_p(X,Z)\propto\sum^{+\infty}_{n=-\infty}C_ne^{-i\frac{\pi\lambda_pn^2Z}{a^2}}e^{i\frac{2\pi{n}X}{a}}.
\end{eqnarray}
Some conclusions are immediately in order from Eq. (5). First, in
the planes $Z_T=ma^2/\lambda_p$, where $m$ denotes a positive
integer referred to as the self-imaging number, the field amplitude
matches the amplitude at the output plane of the ensemble $z=L$. At
these planes, all the diffraction orders are in phase. In the case
of $m$ being odd integers, the self-images are shifted half a period
with respect to the even-number orders. Second, the localization of
the planes with best visibility is independent of the magnitude of
the phase modulation due to the EIT effect. It coincides with the
self-image planes. In general, the axial repetition period
$a^2/\lambda_p$ of the Fresnel diffraction is preserved. Third,
there are no planes where only images of phase modulation occur.

The validity of Eq. (4) can be verified by numerically evaluating
Eq. (3) with use of (2). For instance, in Fig. 3(a) we gave the
spatial distribution of the diffraction amplitude at the first
Talbot plane ($m=1$) by setting $\Delta_p=\Delta_2=0$, i.e., an
amplitude grating. From the figure, it is obvious that the
\textit{object} image is shifted laterally half a period with
respect to the output profile of the probe field [see Fig. 2(a)],
which agrees with the theoretical prediction described above. This
further proves the correctness of Eq. (4) using the Fourier series
expansion, which allows the analytical analysis on the effect. The
visibility at the Talbot planes approaches almost unity in such an
amplitude-grating case. The spatial profile of the diffracted probe
beam at the $m=1$ plane shown in Fig. 3(b) gave another case that by
detuning it off the resonance, an induced hybrid (both amplitude and
phase) grating can be generated and experienced by the input probe
light. One can notice that the location of the Talbot plane
coincides with the amplitude-grating case and is independent of the
introduced phase modulation, except the maximum amplitude contrast
is decreased [compared with Fig. 2(b)]. All of these agree well with
the prediction drawn from Eq. (5).
\begin{figure}[tbp]
\includegraphics[scale=0.35]{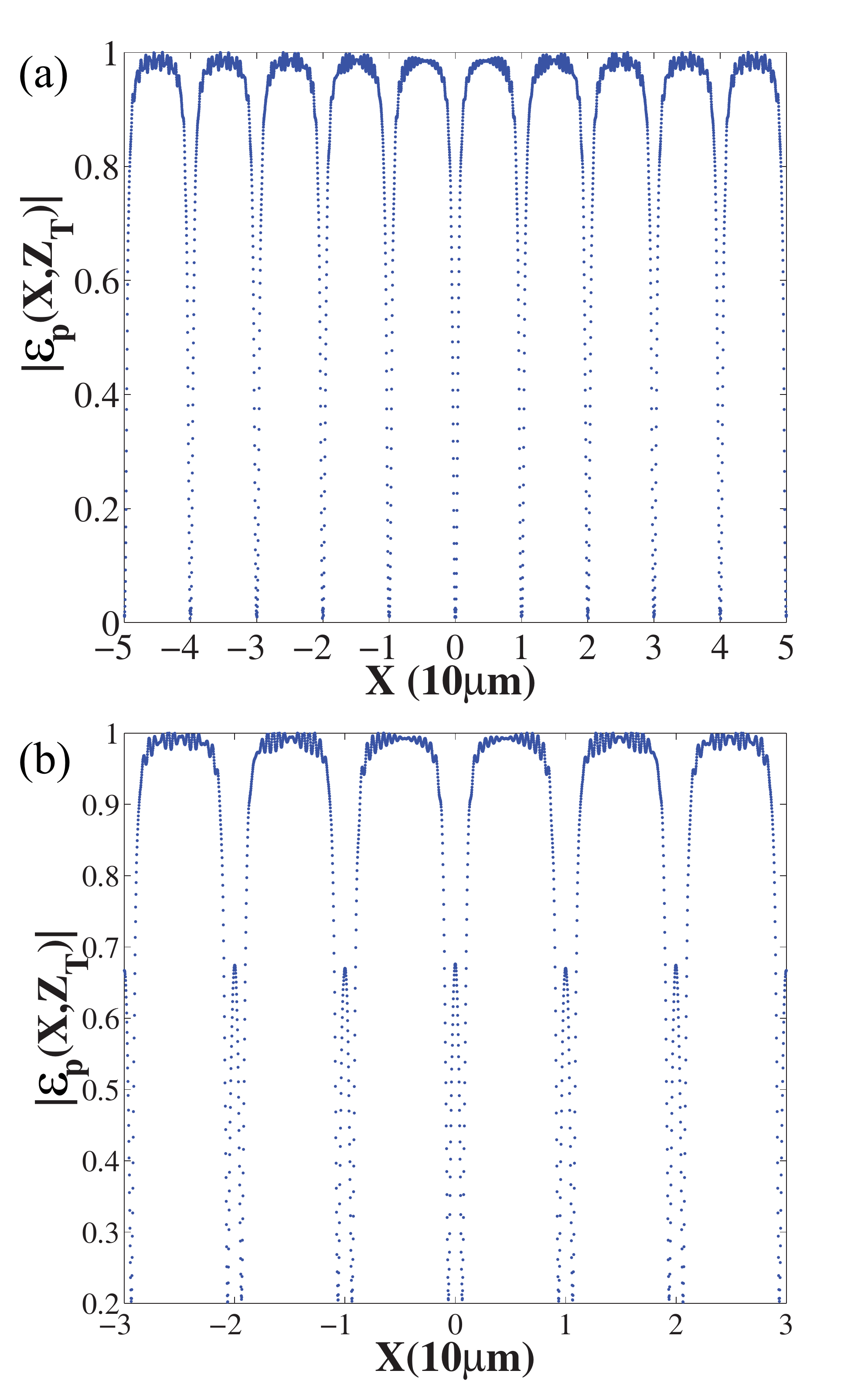}
\caption{Self-imaging of the output probe field,
$|\mathscr{E}_p(X,Z_T)|$, observed at the first Talbot plane
($m=1$). (a) An amplitude grating: parameters are chosen as optical
depth $=20$, $\Omega_c=10\gamma_a$, $\Delta_p=\Delta_2=0$,
$\gamma_{bc}=0.3\gamma_a$, and $a=10$ $\mu$m. (b) A hybrid grating:
optical depth $=8$, $\Omega_c=10\gamma_a$, $\Delta_p=2\gamma_a$,
$\Delta_c=0$, $\gamma_{bc}=0.3\gamma_a$, and $a=10$
$\mu$m.}\label{fig:Fig3}
\end{figure}

In summary, we theoretically propose an idea to image the cold atoms
with EITE. The effect is capable of lensless imaging and may reduce
the influence from vibrations in the experiment. Thus it could be
useful for imaging BEC on chip \cite{hansel,du} and optical lattice.
In practice, the imaging quality may be affected by the finite
dimensions of the standing wave and the size of the input probe
field. For visible light, the Talbot length here is about several
tenth of millimeters. Therefore, a second imaging process would be
necessary to magnify the self-images as implemented in
Ref.\cite{wen2} We also notice that our scheme might be possible to study the nonspreading wave packets as demonstrated in Ref. \cite{stutzle}. Further development of this proposal will be
presented elsewhere. Besides, our recent research \cite{kerr}
indicate that such a configuration is also useful for quantum
information science.

We gratefully acknowledge helpful discussions with M. H. Rubin and
Yan-Hua Zhai. J.W. and M.X. were supported in part by the National
Science Foundation (USA). J.W. also acknowledges financial support
from China by 111 Project B07026. S.D. was supported by the Hong
Kong Research Grants Council (Project No. HKUST600809).

\end{document}